\documentclass[prb,twocolumn]{revtex4}

\usepackage{graphicx}
\usepackage{ifthen}
\usepackage{ifpdf}
\usepackage{color}

\usepackage{latexsym}
\usepackage{amsmath}
\usepackage{amssymb}
\usepackage{bm}
\newcommand{\half}{\mbox{\small $\frac{1}{2}$}}

\newcommand{\eexp}{\mbox{e}^}

\newcommand{\beq}[1]{\begin{eqnarray}\ifthenelse{#1=-1}{\nonumber}
{\ifthenelse{#1=0}{}{\label{e#1}}}}
\newcommand{\eeq}{\end{eqnarray}}
\newcommand{\hide}[1]{}

\begin{document}
\title{Fingerprints of single nuclear spin energy levels using STM - ENDOR }
\author{Yishay Manassen$^1$, Michael Averbukh$^1$, Moamen Jbara$^1$, Bernhard Siebenhofer$^1$,  Alexander Shnirman$^2$ and Baruch Horovitz$^1$}
\affiliation{$^1$Department of Physics and and Ilse Katz Center of Science and Technology in the nm scale, Ben-Gurion University of the Negev, Beer-Sheva 85105, Israel\\
$^2$Institut f\"ur Theorie der Kondensierten Materie, Karlsruhe Institute of Technology,
D-76131 Karlsruhe, Germany}
\begin{abstract}
We performed STM-ENDOR experiments where the intensity of one of the hyperfine components detected in ESR-STM  is recorded while an rf power is irradiated into the tunneling junction and its frequency is swept. When the latter frequency is near a nuclear transition a dip in ESR-STM signal is observed. This experiment was performed in three different systems: near surface SiC vacancies where the electron spin is coupled to a next nearest neighbor $^{29}$Si nucleus;  Cu deposited on Si(111)7x7 surface, where the unpaired electron of the Cu atom is coupled to the Cu nucleus ($^{63}$Cu, $^{65}$Cu) and on Tempo molecules adsorbed on Au(111), where the unpaired electron is coupled to a Nitrogen nucleus ($^{14}$N). While some of the hyperfine values are unresolved in the ESR-STM data due to linewidth we find that they are accurately determined in the STM-ENDOR data
including those from remote nuclei, which are not detected in the ESR-STM spectrum. Furthermore, STM-ENDOR can measure single nuclear Zeeman frequencies,  distinguish between isotopes through their different nuclear magnetic moments and detect quadrupole spectra. We also develop and solve a Bloch type equation for the coupled electron-nuclear system that facilitates interpretation of the data. The improved spectral resolution of STM - ENDOR opens many possibilities for nanometric scale chemical analysis.
\end{abstract}
\maketitle


\section{Introduction}

The attempt to detect and manipulate a single spin is a fundamental challenge in nanoscience and nanotechnology. For that purpose, several low temperature scanning tunneling microscopy (STM) techniques have been developed. In particular an electron spin resonance (ESR) detection has been developed by annalyzing the current power spectrum of an STM, a technique known as ESR-STM \cite{balatsky}. A related technique measures DC spin polarized current in presence of variable rf frequency around the Larmor frequency \cite{baumann,mulleger}.

In this work we develop a novel technique for detection of single electron and nuclear resonance. This is based on ENDOR (Electron Nuclear Double Resonance), i.e. a technique where rf field frequencies are swept across nuclear transitions which are then detected via intensity changes of a simultaneously irradiated ESR (Electron Spin Resonance) transition.  This is possible when there is a coupling between the electron and the nuclear spins. The spin Hamiltonian is then
\beq{01}
{\cal  H}_0=\gamma_eH_0S_z +\gamma_n H_0I_z + {\bf S}\cdot \hat a\cdot {\bf I}
\eeq
 where ${\bf S},\,{\bf I}$ are the electron and nuclear spin operators, respectively, $\gamma_e,\,\gamma_n$ are the corresponding gyromagnetic ratios (e.g. for an electron's g factor of 2 $\gamma_e=2.8$MHz/G and for $^{29}$Si nucleus $\gamma_n=-8.4$MHz/T), $H_0$ is a DC magnetic field in the $z$ direction and $\hat a$ is the hyperfine tensor. The electron and nuclear Zeeman energies are defined as $h\nu_e=\gamma_eH_0$, $h\nu_n=\gamma_nH_0$, respectively.

\begin{figure}  \centering
\includegraphics [width=.5\textwidth]{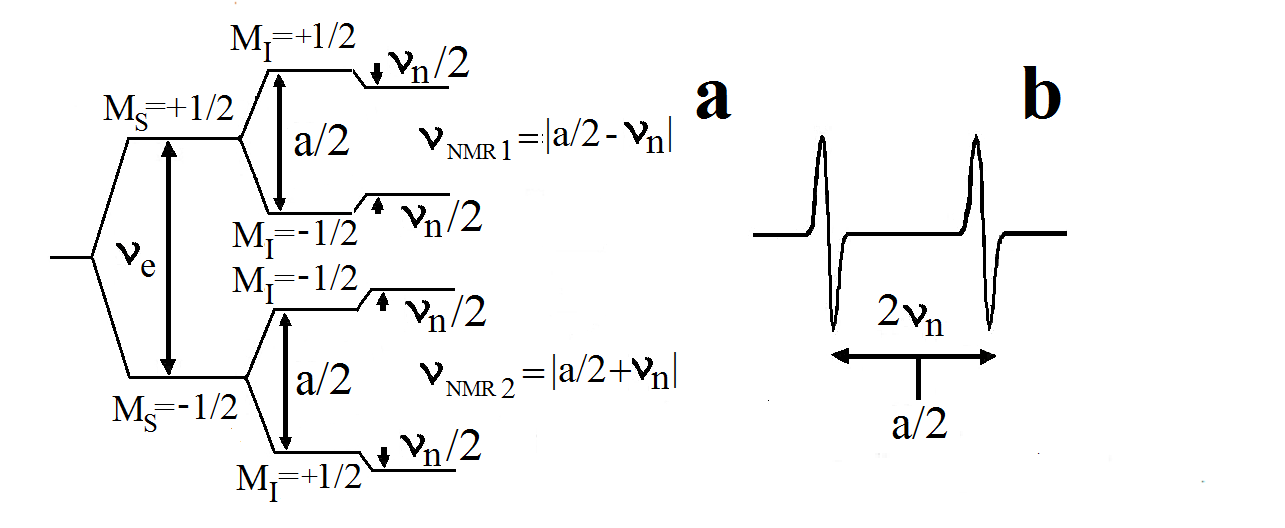}
\caption { (a) Energy level diagram for an electron and nuclear spin $S=\half,\, I=\half$ respectively, illustrating the electron Zeeman frequency $\nu_e=\gamma_eH_0$, nuclear Zeeman frequency $\nu_n=\gamma_nH_0$ and hyperfine splitting for the case where $a=a_z>0$ and $\half a > \nu_n$ ; $M_S=\pm\half$, $M_I=\pm\half$ are the spin projections. The nuclear transitions, labeled as $\nu_{NMR1},\nu_{NMR2}$, (b) The usual ENDOR spectrum – in low magnetic field were $\half a>\nu_n$ gives signals at the two NMR transitions.}
\label{levels}
\end{figure}

 The simplest case $S=I=\half$ is shown \cite{murphy} in Fig. \ref{levels}a. In this case, one has two ESR transitions at $\nu_e\pm \half a$, where a is the component of $\hat a$ parallel to the magnetic field and $a\ll \nu_e$ is assumed, and two nuclear transitions at $|\half a\pm \nu_n|$. Thus, when $\half a>\nu_n$ (as in our low field experiments) the two nuclear transitions are separated by $2\nu_n$, identifying the NMR frequencies Fig. \ref{levels}b. In the usual ENDOR method \cite{feher} one of the ESR transitions is saturated so that the level populations become equal and there is no (or little) absorption. Irradiation at the NMR frequency involves a third state with an opposite nuclear spin and therefore will unequilize the ESR levels populations, hence the  ESR intensity is partially restored \cite{feher,abragam}. A distinct type of "negative ENDOR" \cite{lambe,miyagawa,doan} is obtained by applying a strong rf field that modifies the ESR signal and then the ESR intensity at the original peak is reduced.

   The ability of ENDOR to detect the nuclear  transition frequencies, combined with the ability to detect single electron spins by STM techniques, such as ESR-STM, opens the possibility to detect the nuclear transition frequencies of a single atom, once the hyperfine spectrum is detected. This is the topic of this paper. The technique of ESR-STM is capable of detecting single spins. In this method, a Larmor frequency component of the tunneling current is induced by the precession of a nearby single spin on the surface. The existence of this phenomenon has been demonstrated on several spin systems, allowing also observation of hyperfine coupling \cite{manassen1,manassen2,manassen3,manassen4,durkan1,durkan2,
   naruszewicz,morgenstern,komeda1,komeda2,balatsky}. We note that the theoretical understanding of the phenomena seen in ESR-STM is a subject of  ongoing research \cite{balatsky,bh}.
   The more recent proposal \cite{bh} employs spin-orbit coupling as well as an additional direct current path from the tip to the substrate. Detailed calculations show that the interference between the two paths, i.e. the one via the spin and the direct path, produce a Larmor resonance in the power spectrum of the current, i.e. an ESR-STM effect \cite{bh}.

In this work we demonstrate for the first time the feasibility and efficiency of ENDOR within the ESR-STM method. This experiment involves a single external AC field at the nuclear transitions, while the ESR signal is measured by the STM current noise, i.e. its power spectrum. We show negative ENDOR phenomena in a variety of systems: SiC vacancies, Cu on Si(111)7x7 surface and Tempo molecules on Au(111) surface. We also develop and solve a Lindblad equation (equivalent to Bloch's equation) for the coupled electron nuclei system. We show that the ENDOR spectrum, if the AC field is not too strong, gives fairly accurate values of the hyperfine coupling and the nuclear Zeeman frequency.

\section{Experimental Methods}

The experimental setup and the magnetic field measurement are described in appendix A (for further details see Ref. \onlinecite{morgenstern}). The setup is modified by assembling a power combiner to add a time dependent AC voltage to the DC tip – sample bias voltage.  The frequency of the AC voltage is slowly swept over the nuclear frequencies of interest, while the spectrum analyzer is recording the intensity of the ESR at a single frequency as function of the (time dependent) AC frequency.  The DC magnetic field is 210G, it has an added small parallel field modulation, and the modulated output of the spectrum analyzer is put in a phase sensitive detector.  Our rf generator has a power of -10dBm, with a tip geometric capacitance of $5\cdot 10^{-13}$F \cite{kurokawa} and a frequency of 10MHz,  taking into account the Bio Savart law and the impedance mismatch with the STM tip we estimate the intensity of the rf field as 1 Gauss or more.
 The coupling of this field to the nuclear spin is enhanced by the mixing of nuclear and electron spins due to $a_\perp$, the component of the hyperfine tensor perpendicular to the field \cite{abragam}. We estimate below that for $^{29}$Si it leads to an enhancement of $\approx 35$ relative to the direct coupling to the nuclear magnetic moment. We note that all the data presented here is an average of 100 sweeps, 1 minute each, over different sites of the sample.

 \begin{figure}  \centering
\includegraphics [width=.4\textwidth]{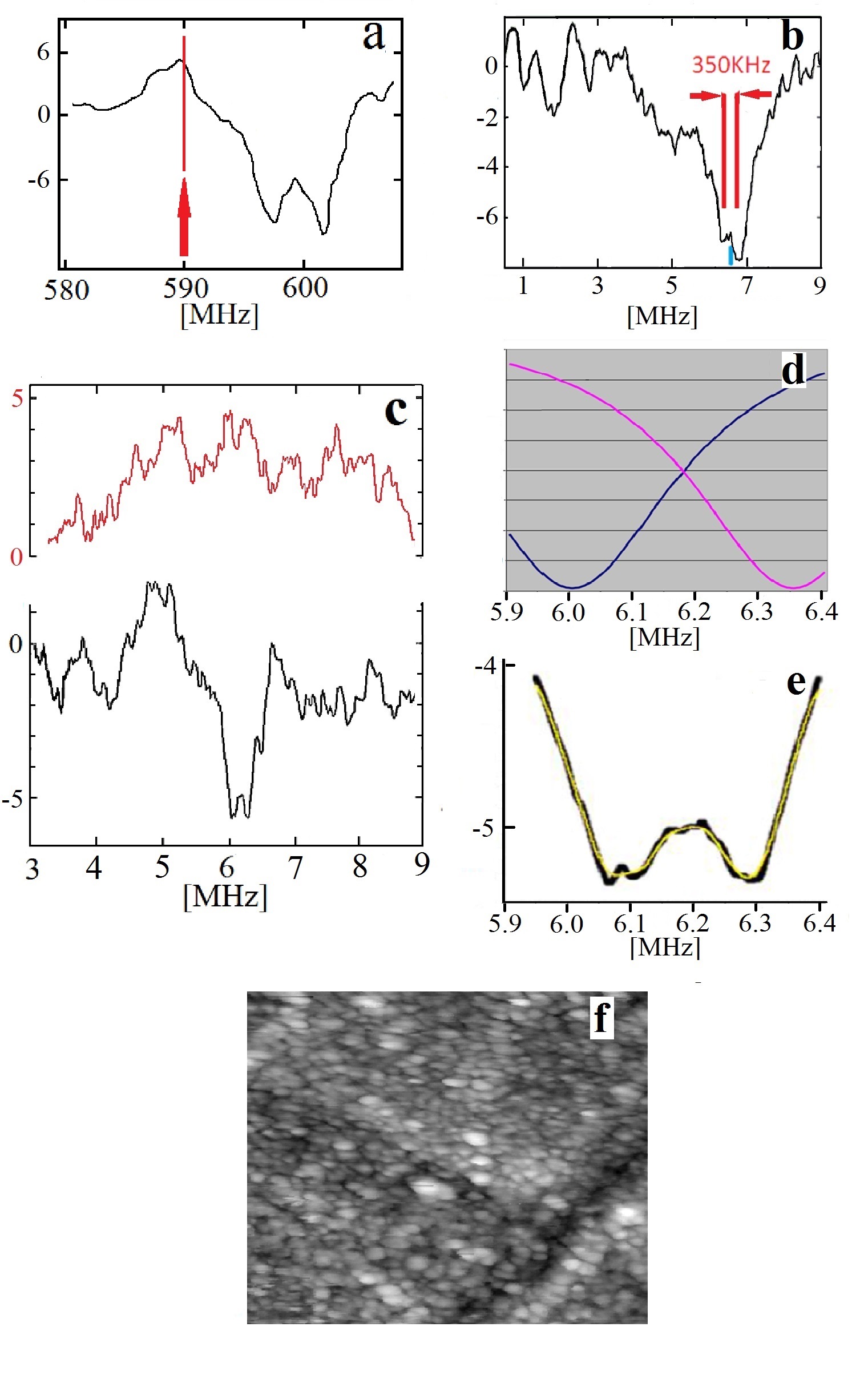}
\caption {(a) ESR-STM data, i.e. the current power spectrum, at 210G. The red arrow points to the monitored hyperfine component at 590MHz. (b) ENDOR spectrum: the intensity at 590MHz as a function of the irradiation frequency showing $\half a$  (blue tick) and $2\nu_n$ splitting (in red). The bandwidth was 300KHz. (c) Similar to (b) except for using a narrower bandwidth of 100KHz.
 The upper (red) curve shows the background Au data. The power spectrum is in units of 10$^{-27}$A$^2$/Hz. (d) Two Lorentzians separated by 350KHz, each of width 100KHz. (e) The sum of the two Lorentzians in (d) (yellow) superimposed on the experimental spectrum from (c) shows the quality of the fitting. (f) Atomically resolved image of 15nm$\times$20nm of disordered SiC observed by the STM while the ENDOR equipment was on. This confirms that no deterioration in the image quality is caused by data acquisition. All measurements here used tunneling current $I_t=0.1$nA and voltage $V_b=+3$V.}
\label{Si}
\end{figure}

\section{Experimental Results}

Fig. \ref{Si}a shows our ESR-STM data on SiC on Si substrate, similar to data in Ref. \onlinecite{morgenstern}. We focus on a line at 590MHz (red arrow) that is an ESR transition shifted by a hyperfine coupling to a $^{29}$Si nucleus. (The presence of $^{28}Si$ with no nuclear spin leads to additional structure, not of interest here \cite{morgenstern}).
Fig. \ref{Si}b shows the change in intensity of the 590MHz line as a result of irradiation of -10dBm of rf with variable frequency between 0.5 and 9MHz. We find a strong negative ENDOR signal centered at 6.4MHz which is split by close to $2\nu_n=354$KHz.

  The bandwidth controlling the 590MHz line in Fig. 2b is 300KHz for the purpose of gaining sensitivity, though this deteriorates the frequency resolution.  Improved results with a bandwidth of 100KHz are shown in Fig. 2c. The shown scale for the power spectrum is in units of 10$^{-27}$A$^2$/Hz. The upper (red) curve shows the background Au data. We note that the rf irradiation at a frequency range as needed in this experiment may cause a non flat background, seen also in the Au background data.

  The center position of the ENDOR dip is at $\half a=6.2$MHz and is clearly absent in the background data. The splitting $2\nu_n$ as shown in Fig. 2b is reproduced also in this setup, although it is within noise level. Fitting two Lorentzians (Fig.2d,e) to the ENDOR data gives indeed a splitting of 350 KHz, in agreement with $2\nu_n$.

  Fig. 2f shows the atomically resolved image which has some signatures of a 7x7 unit cells, altogether, the surface is disordered. Our experience is that this is required so as to localize the wave function of the spin center, which seems to be necessary to observe both ESR-STM and STM Endor signals on SiC surface.

Our data is thus consistent with the negative ENDOR observed in macroscopic samples \cite{lambe,miyagawa,doan}, where the dip intensity was shown to increase strongly with $H_1$,  reaching 10\%. In our data, by comparing with the background noise level \cite{morgenstern}, we estimate the reduction of the ESR-STM signal to be even larger.

\begin{figure}  \centering
\includegraphics [width=.4\textwidth]{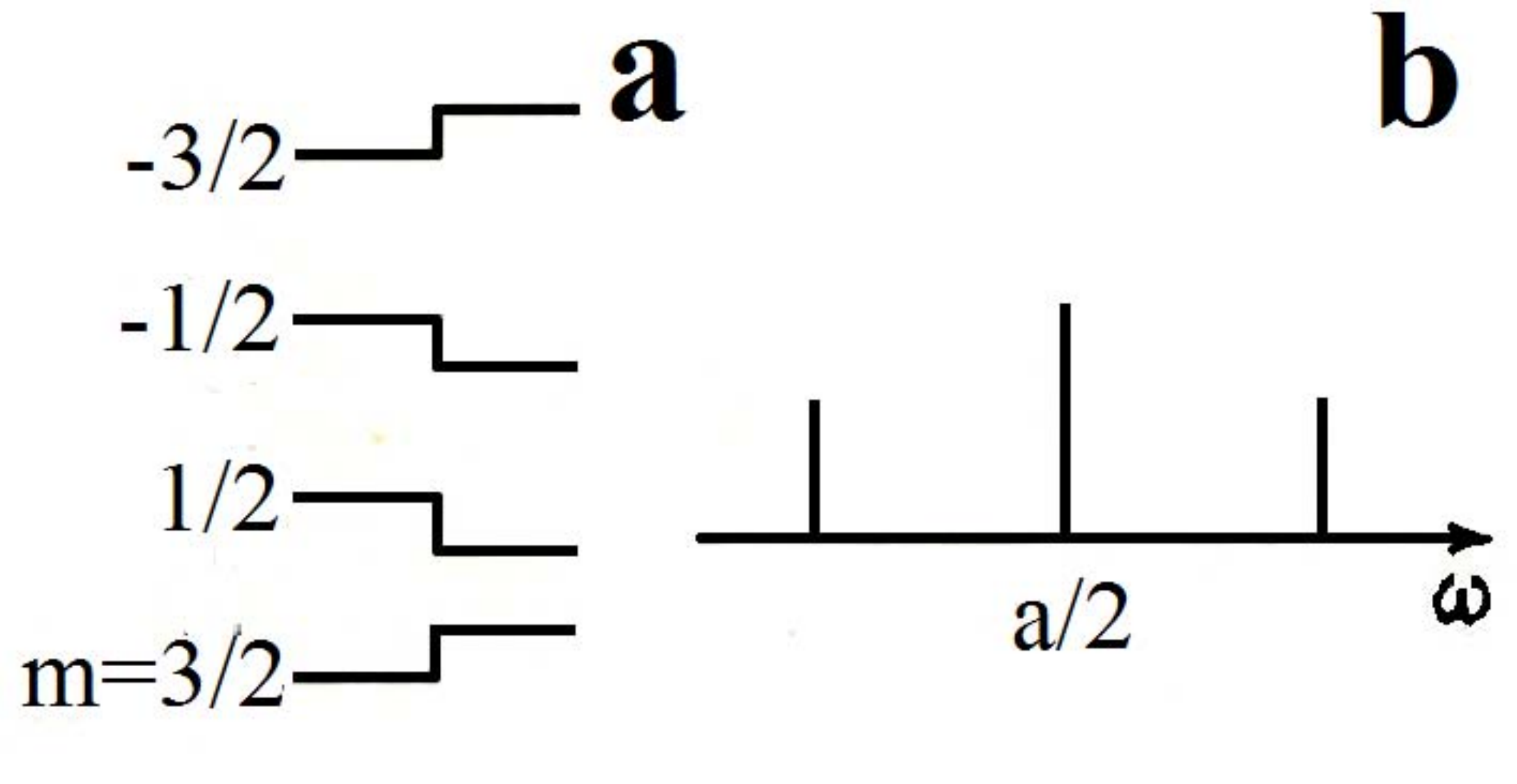}
\caption { (a) Energy levels of a nucleus with $I=3/2$. The left part shows equal spacing $\half a$ (more generally $\half a\pm \gamma_nH_0$) while the right shows the effect of quadruple coupling. The shifts of all levels have the same magnitude. (b) Nuclear transitions of (a) that may be seen in ENDOR.}
\label{3lines}
\end{figure}

To show the general applicability of this technique, we have performed additional experiments on Cu$^{2+}$ nuclei. The ESR spectrum of Copper, is very anisotropic, with an electronic structure of d$^9$ ($S=\half$). The g values are very sensitive to the chemical environment of the Cu atom, with an average of $g_\perp=2.05$ and $g_\parallel=2.23$. The Cu hyperfine tensor is equally anisotropic: $a_\perp = 50-100$MHz and $a_\parallel = 650$MHz \cite{kivelson,finazzo} (the symbols $\perp,\parallel$ refer to the molecular complex plane).  Cu has two isotopes: $^{63}$Cu and $^{65}$Cu with natural abundances of 69\% and 31\%, respectively. The gyro magnetic ratio of $^{63}$Cu is  11.28MHz/T while that of $^{65}$Cu is 12.09MHz/T (larger by 6.7\%), thus a similar relative difference is expected in their hyperfine couplings. Both nuclei have I=3/2 for which a quadruple interaction is expected leading to a triplet spectrum (Fig. \ref{3lines}) \cite{slichter,brown}. Thus, the expected ENDOR lines for each Cu isotope are centered at the corresponding $\half a$ with a quadrupole splitting \cite{brown}; the latter are similar for the two isotopes \cite{gartner}.

\begin{figure}  \centering
\includegraphics [width=.4\textwidth]{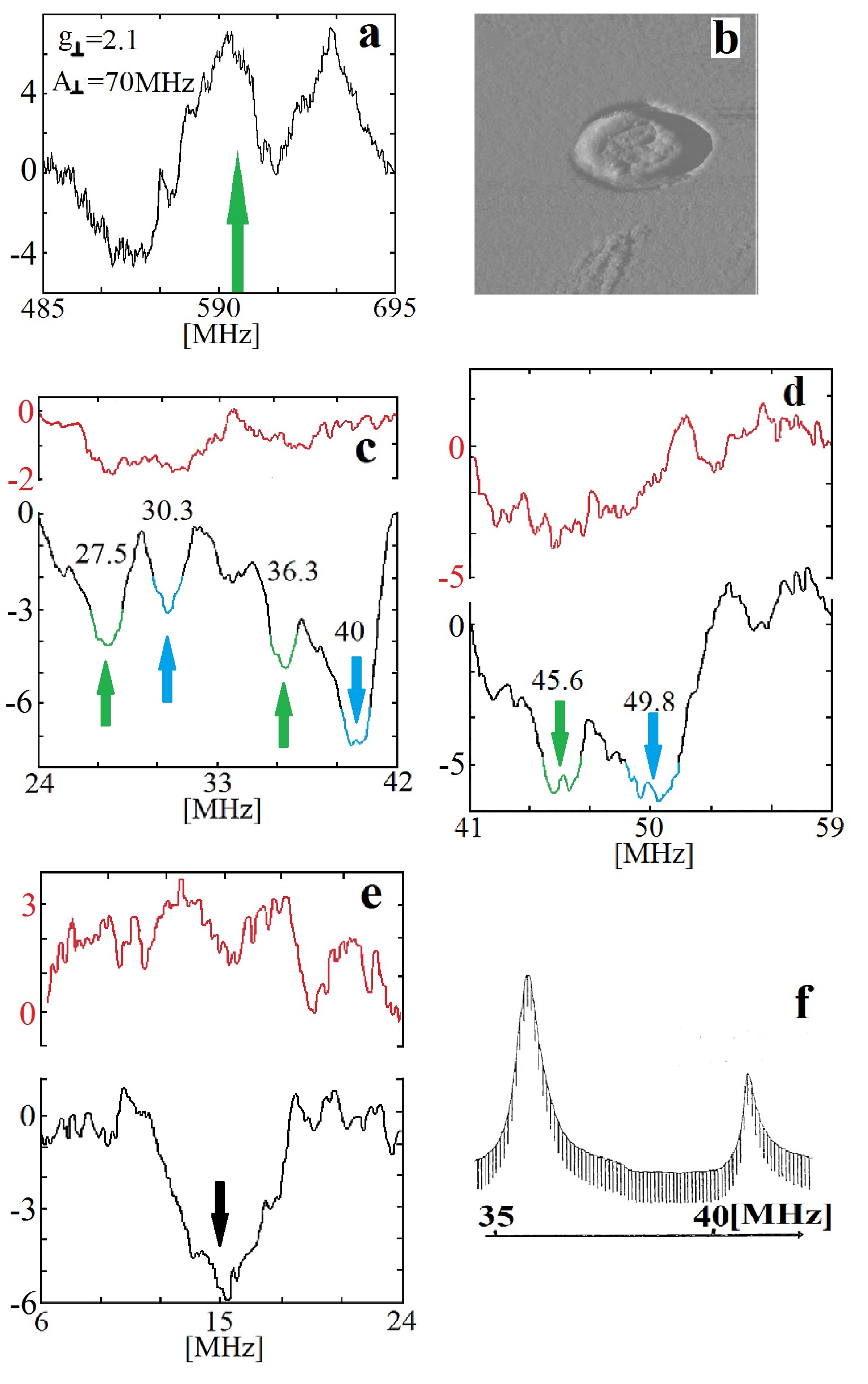}
\caption {  (a) ESR-STM spectrum of one monolayer of Cu on Si(111)7x7 at a magnetic field of 212G. (b) :  hole – island pair of Cu on Si(111)7x7. 50pA tunneling current; Bias voltage 2V; scan size 100nm$\times$100nm.   (c),(d) STM ENDOR of this sample observed in two different ranges, here $I_t= 0.2$nA and $V_b$ = -3V, detector bandwidth is 30KHz. The green and blue colors in the spectra correspond to the triplets of the two distinct Cu isotopes. Here and in (e) the upper (red) curves are for the Au substrate.  The power spectra are in units of 10$^{-27}$A$^2$/Hz. (e) STM ENDOR of Cu on Si(111)7x7 at a magnetic field of 212G – at lower frequencies – presumably hydrogen frequencies. $I_t$ = 0.2 nA, $V_b$ = -1V, detector bandwidth 30KHz. (f) For comparison a partial spectra (two peaks) from macroscopic ESR showing quadrupole interaction of $^{63}$Cu, taken from Fig. 6 of Ref. \onlinecite{schweiger}.}
\label{Cu}
\end{figure}

The experiment was done by thermal evaporation of 1 monolayer of Cu on clean Si(111)7x7 surface. During the evaporation the sample was held at room temperature. Afterwards, the sample was heated for 10 minutes to 500$^{\circ}$C. The STM image (Fig. 4b) shows a typical structure of such a surface \cite{tosch,yasue,saranin}.
In this STM system, the magnetic field is oriented parallel (perpendicular) to the tip (sample). Thus the structure of the Cu atom on the surface is  reminiscent of the situation in flat complexes of Cu \cite{finazzo}, therefore the $g_\perp$ and  $a_\perp$ values are relevant to us. Fig. \ref{Cu}a shows our ESR-STM data on this system, showing 3 lines, (the expected 4th one is outside our range)  consistent with the known $g_\perp,\,a_\perp$ values.
The ENDOR spectrum is measured by monitoring the intensity of the 600MHz peak (marked with a green arrow in Fig. \ref{Cu}a) while a rf generator with -10dBm is swept in frequency in the ranges shown in Figs. \ref{Cu}c,d. The results are remarkable in that they show two triplets of lines, confirming our success of observing Cu ENDOR. We associate the green triplet in Figs, \ref{Cu}c,d with $^{63}$Cu while the blue triplet with $^{65}$Cu. The central peaks of these triplets at 36.3MHz and 40MHz, respectively, differ by 9.2\%, close to the expected difference in $a_\perp$ of the two isotopes. The central peak positions are also close to $\half a_\perp$ as measured by ESR-STM (Fig. \ref{Cu}a) The splittings within each triplet are similar, in the range of 9-10MHz for $^{65}$Cu, consistent with the known quadruple moments. The splitting is similar, though somewhat larger than that observed in macroscopic ENDOR (Fig. 6 of Ref. \onlinecite{schweiger}, shown in Fig. \ref{Cu}f), and with other data \cite{hyde}, the difference could be due to variations in local electric field gradients \cite{wesche}

We have done an additional experiment at lower ENDOR frequencies (Fig. \ref{Cu}e). These frequencies are presumably due to protons. We have observed a strong signal at 15MHz. This signal is known in other Cu systems to be related to unbound hydrogen \cite{roberts,whittaker}. We did not put intentionally hydrogen into the UHV system, but it is known to be a present in relatively large amount as a residual gas.

\begin{figure}  \centering
\includegraphics [width=.4\textwidth]{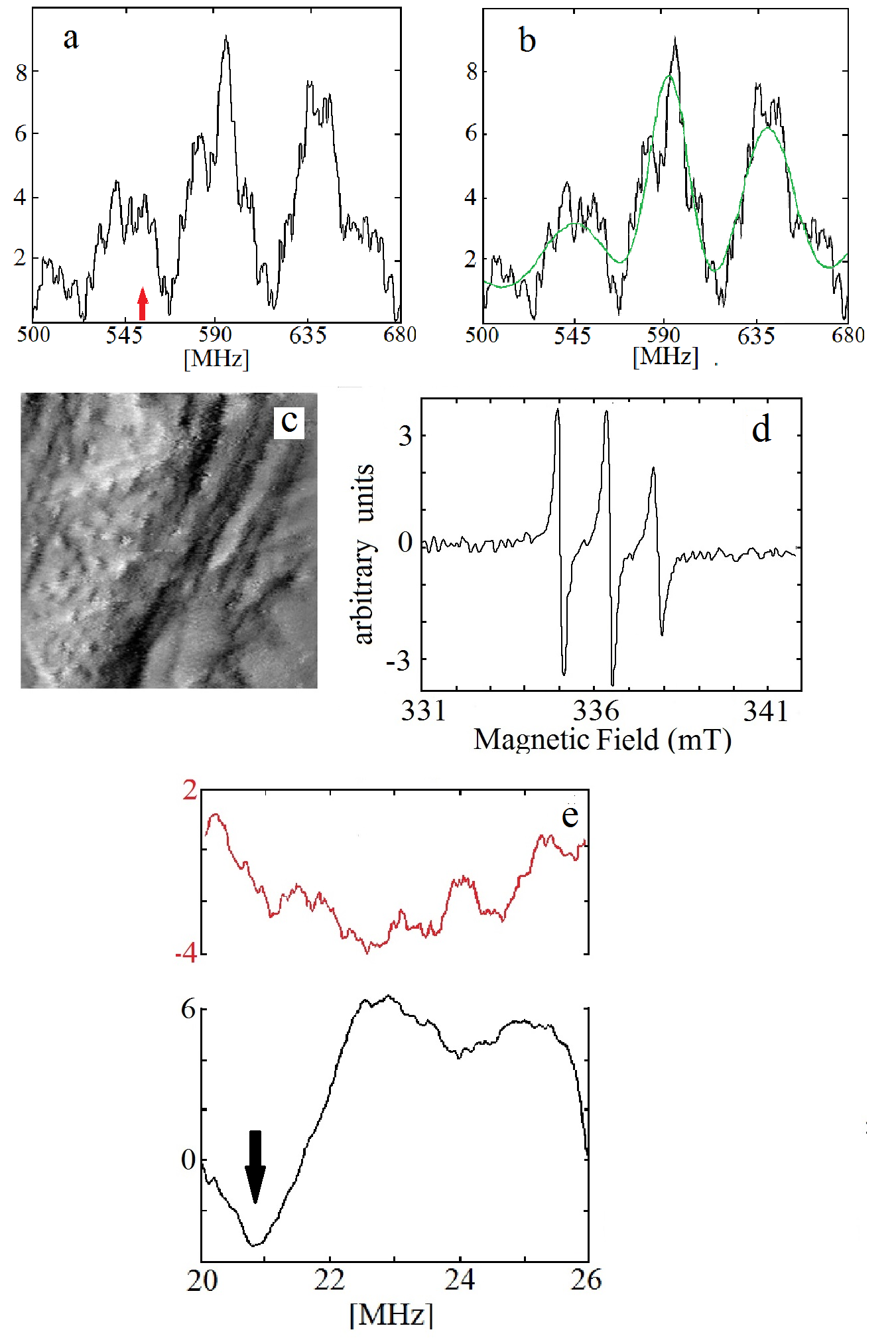}
\caption{(a) ESR-STM spectrum of Tempo on Au(111) surface, (b) same data with statistical analysis. The data shows hyperfine splitting, in agreement with the macroscopic ESR spectrum shown in (d). The arrow in (a) is pointing to the ESR frequency in which the ENDOR signal was collected as a function of the AC field frequency.  (c) STM image 70nm$\times$70nm of Tempo on Au(111), $I_t=0.2$nA, $V_b=200$mV. (e) The ENDOR spectrum of TEMPO measured as the intensity change at 557.5MHz (the arrow in (a)). The arrow here shows a single ENDOR frequency at 21MHz, close to the expected half of the hyperfine splitting in (a). The upper (red) curves is for the Au substrate.  The power spectra are in units of 10$^{-27}$A$^2$/Hz.  }
\label{tempo}
\end{figure}

We performed further experiments on TEMPO molecule on Au(111) substrate.  In this case the ENDOR transitions are due to $^{14}$N and $^1$H nuclei. The hyperfine spectrum of Tempo on Au(111) was observed with ESR-STM measurement on TEMPO, Fig. \ref{tempo}a, which is in clear agreement with the macroscopic ESR spectrum of TEMPO, Fig. \ref{tempo}d. We show our ENDOR data in Fig. \ref{tempo}e, focusing on the $^{14}$N range. We infer a hyperfine coupling, i.e. twice the arrow position in Fig. \ref{tempo}e, of 42MHz, while the ESR-STM measurement, Fig. \ref{tempo}a, shows a 45MHz splitting.  We note that the macroscopic ENDOR data \cite{kotake1} of similar molecules shows a hyperfine of 48MHz as well as nuclear Zeeman splitting. The latter are below our resolution in our weaker magnetic field; $^{29}$Si with its much larger gyromagnetic ratio allows our observation of Nuclear Zeeman, Fig. \ref{Si}b,c.

We also show statistical analysis of the ESR-STM data in Fig. \ref{tempo}b, using the EasySpin software \cite{stoll}. The simulation applied a Monte Carlo fitting algorithm to observe a spectrum that fits the observed experimental result (green line in Fig. \ref{tempo}b). The fit gives a root mean square deviation of 9.6\% at different initial values. Moreover, the simulation was able to give correct values of the g and the hyperfine tensors (in the principle coordinate system), with reasonable accuracy (parentheses give the relative error from published values): $g_x$ = 2.00243 (0.37\%), $g_y$ = 2.03948 (1.6\%), $g_z$ =1.95915 (2.16\%); $A_x$ = 16.4129 MHz (2.39\%), $A_y$ = 15.0084 MHz (2.63\%), $A_z$ = 97.0728  MHz (3.4\%).

\section{Theory}

We proceed to describe our theoretical approach, aiming to solve the time dependent problem of our ENDOR experiment and examining the conditions for observing the nuclear Zeeman frequency. We wish to solve the Hamiltonian of Eq. (\ref{e01}) when an AC field is added as well as a coupling to an environment that leads to relaxation and dephasing. The full Hamiltonian is then ($\omega_e=2\pi\nu_e,\,\omega_n=2\pi\nu_n$)
\beq{02}
{\cal H}/\hbar&=&\omega_eS_z+\omega_nI_z+aS_zI_z+\gamma_eH_1S_x\cos\omega t\nonumber\\&+&a_\perp(S_xI_x+S_yI_y)+
X S_++X^*S_- +ZS_z
\eeq
where $X,Z$ are random Gaussian fields, representing the environment, and $S_\pm=S_x\pm iS_y,\,I_\pm=I_x\pm iI_y$. The additional coupling of $H_1$ directly to the nucleus, $\gamma_nH_1S_x\cos\omega t$ is extremely small and is neglected. The dominant coupling of $H_1$ to the nucleus is due to the mixing term $a_\perp$ \cite{abragam}. We define a unitary transformation of the Schrieffer-Wolff type \cite{SW}, $\eexp{iR}$ where $R=\frac{a_\perp}{\omega_0}(S_yI_x-S_xI_y)$, so that to leading order we obtain
for $\tilde{\cal H}=\eexp{iR}{\cal H}\eexp{-iR}$
\beq{03}
&&\tilde{\cal H}/\hbar=\omega_eS_z+\nu_nI_z+aS_zI_z+4hS_zI_x\cos\omega t \nonumber\\&&+X S_++X^*S_- +ZS_z+O[(a,\nu_n,X,Z)\frac{a_\perp}{\omega_e}]
\eeq
where $h=\gamma_e\frac{a_\perp}{4\omega_e}H_1$ is an effective field that couples between the nuclear states. This coupling represents an enhancement over the direct coupling by a factor of $\frac{\gamma_e}{\gamma_n}\frac{a_\perp}{2\nu_e}\approx 35$ for the $^{29}$Si parameters. This enhancement can be also derived by a classical argument \cite{abragam}: The electron spin follows adiabatically ($\omega\ll \nu_e$) the magnetic field that is tilted from $z$ by an angle $\tan\alpha=\frac{H_1}{H_0}=\frac{\gamma_e H_1}{\omega_0}$. Hence an hyperfine term $a_\perp S_xI_x=a_\perp I_xS_z\frac{\gamma_e H_1}{\omega_e}=\pm \gamma_e\frac{a_\perp}{2\omega_e}H_1I_x$ produces the required nuclear spin coupling.

The negative ENDOR problem is exactly solvable via a Lindblad equation since we have a single strong AC field and the response at the ESR transition can be evaluated by linear response, or as we do by the regression theorem. In fact we need to evaluate the spin correlation that is measured in the ESR-STM experiment rather than the absorption that is measured in usual ENDOR. We proceed then to a rotating frame by the transformation $U=\eexp{2i\omega S_zI_zt}$ which results in a static hamiltonian assuming that we are close to a nuclear resonance, i.e.
$2\omega\gg |\half a-\omega\pm \nu_n|$. It is then straightforward to develop a Lindblad equation that includes relaxation rates $\Gamma_r$ ($\Gamma_e$) from the electron excited (ground) state as well as a dephasing rate $\Gamma_\phi$; details are given in appendix B.

We present here the spin correlation function for $S=I=\half$ and for three values of $h$, corresponding to $H_1=4,8,12G$ and other parameters corresponding to the Si case, showing negative ENDOR. The ESR frequency is slightly detuned from resonance, showing an assymetry as in Fig. \ref{Si}b. At resonance $\nu=590MHz$ the curves are symmetric. We note that the distance between the dips is close to $2\nu_n$, though increasing with $H_1$. The peak in between the dips is close to $\half a$, very weakly dependent on $H_1$ (see Eq. (5) of the Supplementary material). We note that the chosen relaxation rates yield a line width of $\approx 0.5MHz$ for the ESR line; the observed linewidth is somewhat larger, probably reflecting inhomogeneous broadening due to averaging on sites.

\begin{figure}  \centering
\includegraphics [width=.4\textwidth]{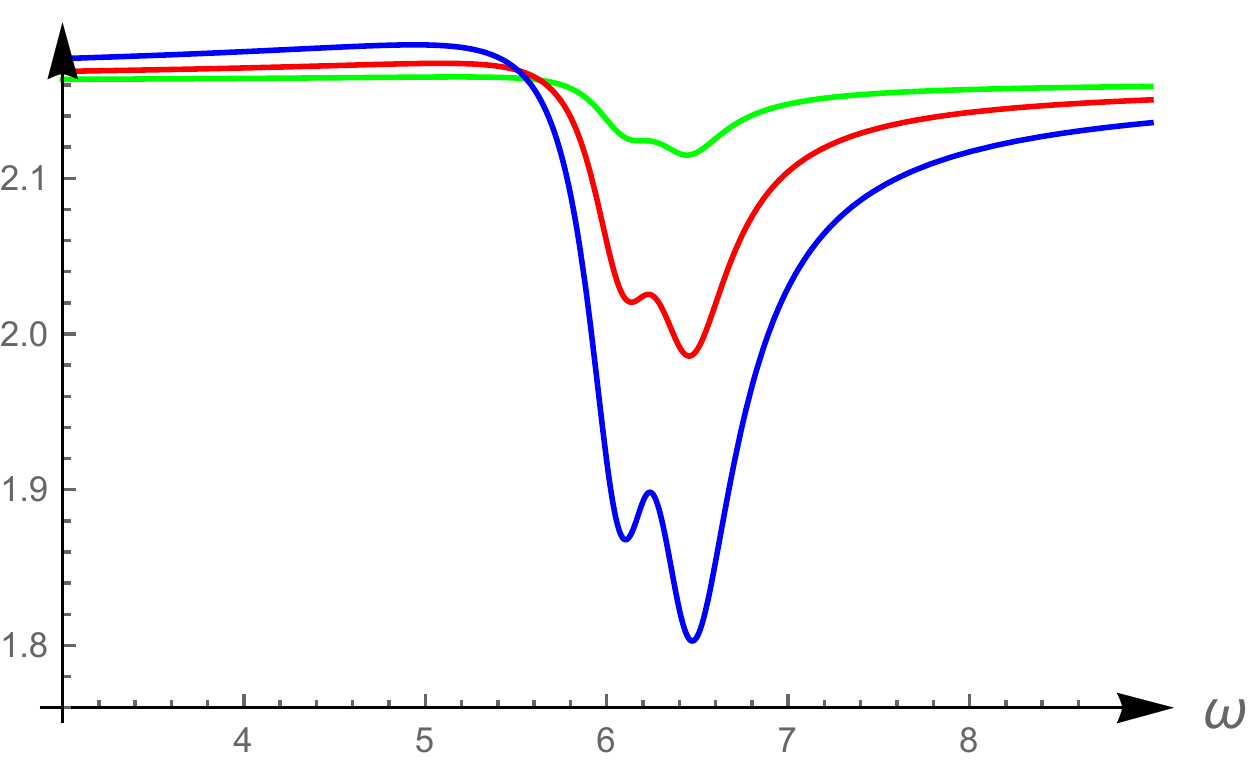}
\caption{Spin correlation Fourier transformed $\langle S_-(t)S_+(0)\rangle_\omega$ corresponding to
$H_1=4,8,12G$, from top to bottom lines. The parameters are, in MHz units: $\nu_e=596.2,\,a=12.4, \nu=589.95, \nu_n=0.17, \Gamma_e=\Gamma_r=0.2, \Gamma_\phi=0.02$.}
\label{regression}
\end{figure}

\section{Conclusions}

In conclusion, we have shown that by irradiating the nuclear energy levels and following the changes in the hyperfine peaks observed in the ESR-STM measurement it is possible to observe detailed chemical information of the single spins under the tip. This includes hyperfine couplings, ligand hyperfine couplings, quadrupole splitting and nuclear Zeeman transitions. The technique was demonstrated to work in three types of spins (defects, adsorbed metal atoms and paramagnetic molecules). Our theoretical achievement for negative ENDOR facilitates interpretation of the data.
Nanometric scale chemical analysis with improved spatial and spectral resolutions is now possible.

\begin{acknowledgments}
This work was funded by the Marie Curie grant from the European commission. Grants from the
the Israel Science Foundation (Bikura and Regular ISF programs), from the ministry of science infrastructure program and from the DFG project "Magnetism of vacancies and edge states in graphene probed by electron spin resonance and scanning tunneling spectroscopy" are gratefully acknowledged. AS acknowledges support from the DFG Research Grant SH 81/3-1.
\end{acknowledgments}

\appendix

\section{STM setup}

\begin{figure}  \centering
\includegraphics [width=.5\textwidth]{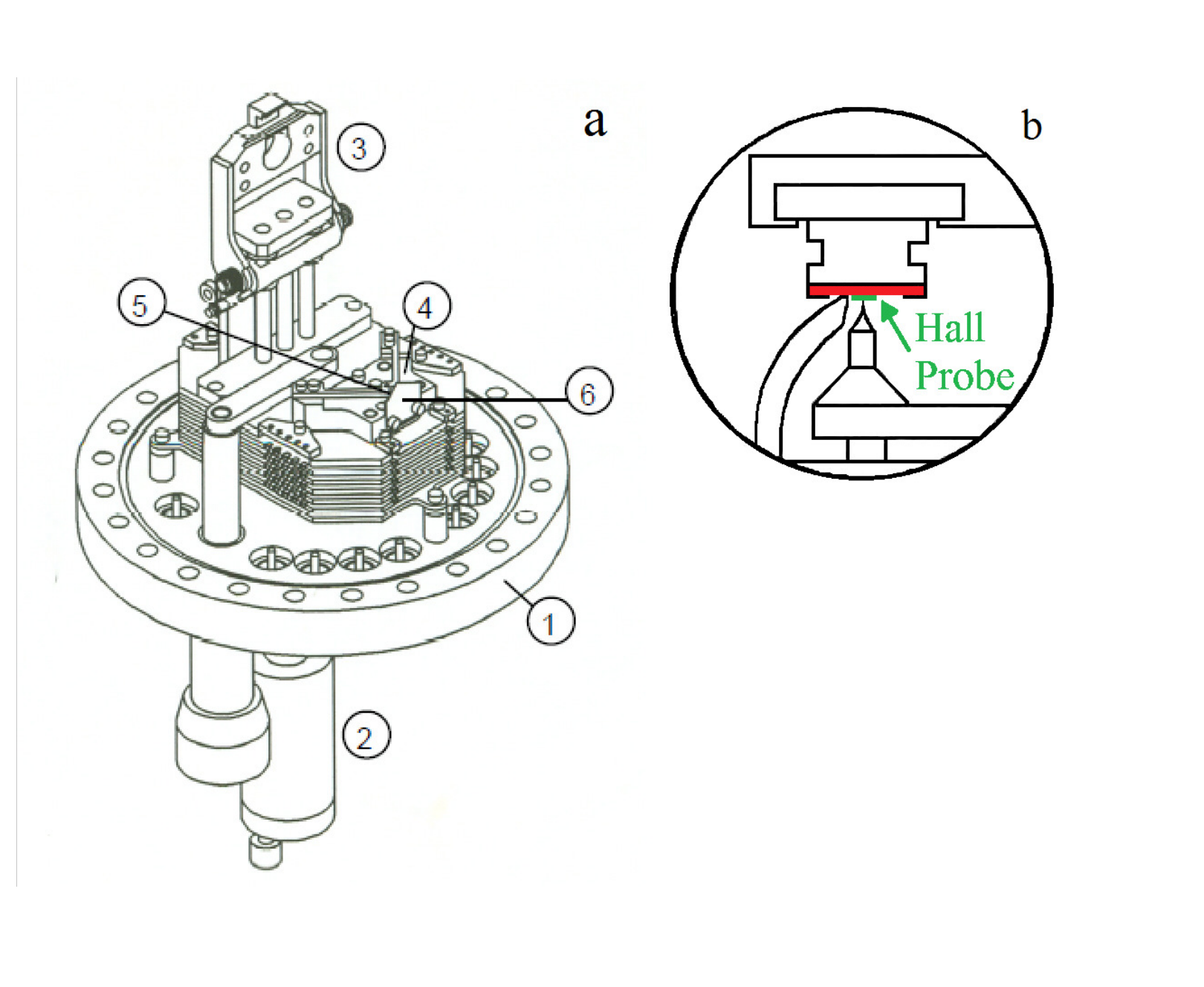}
\caption{(a) The STM setup:  (1) The 8" base flange. (2) The mechanical approach micrometer. (3) Sample holder on coarse approach lever. (4) Piezoelectric tripod for the tip. (5) Tip position. (6) Sample holder support (foot). (b) The Hall probe (in green) attached to the sample holder (in red) at a position close  to the tip, measuring the magnetic field.}
\label{STM}
\end{figure}

The Demuth type STM \cite{julian}, as shown in Fig. \ref{STM}a, is constructed on 8" CF flange that can be mounted onto a UHV chamber. The tip is mounted on a piezoelectric tripod for xy scanning and tip-sample (z) separation control. The sample can be replaced in situ and it can be mounted on a lever that by a micrometer screw bends to serve as a coarse approach mechanism. After the lever is brought down, such that the sample holder (with either the sample or the Hall probe) rests on a supporting bench (the foot), an additional turning of the micrometer screw twists the lever for a fine tuning of the mechanical approach – until tunneling is reached.
    
 Two bar magnets (not shown in Fig. \ref{STM}) were loaded in the STM. Such magnets create a homogenous magnetic field over a volume of several mm$^3$, giving a precise knowledge of the field within $\pm 0.1$G (we estimate this accuracy using different ESR-STM runs with the same tip).
The magnetic field is measured with the Hall probe (green in Fig. \ref{STM}b) replacing precisely the sample position on the sample holder (red in Fig. \ref{STM}b). A foot is aligned in a small lateral distance from the tip (see figure \ref{STM}b). When the Hall probe is approached to the tip, the sample holder first touches the foot and only afterwards by further approach it reaches close to the tip. Using an optical microscope, we can move the Hall probe to a distance of few microns just above the edge of the tip. In this way we estimate the accuracy of the measurement as $\pm 1$G. This accuracy can be improved afterwards using the ESR-STM data.

\section{Theory Methods}

We describe here the derivation and solution of the Lindblad equation for the problem of negative ENDOR. We start from the Hamiltonian $\tilde{\cal H}$, Eq. (\ref{e03}) of the main text, and transform to a moving frame using $U=\eexp{2i\omega S_zI_zt}$
\beq{11}
\tilde{\cal H}_{rot}/\hbar=\omega_0S_z+\omega_nI_z+(a-2\omega)S_zI_z+2hS_zI_x+\tilde{\cal H}_{SE,rot}\nonumber\\
\eeq
 The counter-rotating term $2hS_z I_+\eexp{4i\omega S_zt}+h.c.$ is neglected, valid for $2\omega\gg|\half a-\omega\pm\omega_n|$.
 The system-environment (SE) interaction involves
 \beq{12}
 &&\eexp{2i\omega S_zI_zt}S_+\eexp{-2i\omega S_zI_zt}=S_{a+}\eexp{i\omega t}+S_{b+}\eexp{-i\omega t}\nonumber\\
 && S_{a+}=S_+(\half+I_z),\qquad S_{b+}=S_+(\half-I_z)\nonumber\\
 && \tilde{\cal H}_{SE,rot}=[X(S_{a+}\eexp{i\omega t}+S_{b+}\eexp{-i\omega t})+h.c.]+ZS_z\nonumber\\
 \eeq
 To find Lindbald's equation \cite{gardiner}, we neeed, in principle, to find ${\cal H}_{SE}$ in the interaction picture with respect to system terms in $\tilde{\cal H}_{rot}$. Since $\omega_0$ is by far the largest frequency we can use only $\eexp{-i\omega_0S_z t}$ to define the Lindblad equation and then return to the frame of Eq. (\ref{e11}) that generates $-i[\tilde{\cal H}_{rot},\rho]$ below. Lindblad's equation for the system's reduced density matrix $\rho$ (i.e. after integrating the environment) is identified by the correlations of $X, Z$ and the operators
 \beq{13}
 &&A(t)=\sum_j    A_j\eexp{-i\nu_j t}\qquad  A_0=S_z,\,\,\,\, \nu_0\approx 0 \nonumber\\
&& A_{\pm 1}=S_{a\pm},\,\,\,\, \nu_{\pm 1}=\pm(-\omega_0-\omega)\nonumber\\
 && A_{\pm 2}=S_{b\pm},\,\,\,\, \nu_{\pm 2}=\pm(-\omega_0+\omega)\nonumber\\
 &&\gamma(\pm\omega_0)=2\int_0^\infty dt\, \eexp{\mp i\omega_0 t}\langle X(t)X^*(0)\rangle,\nonumber\\ && \gamma(0)=2\int_0^\infty dt\, \langle Z(t)Z(0)\rangle\nonumber\\
 &&\frac{d}{dt}\rho=-i[\tilde{\cal H}_{rot},\rho]+\nonumber\\ &&\qquad \sum_j\gamma(\nu_j)[A_j\rho A_j^\dagger-\half A_j^\dagger A_j\rho -\half \rho A_j^\dagger A_j]
 \eeq
 where $-i[\tilde{\cal H}_{rot},\rho]$ appears since we returned to the rotating frame of Eq. (\ref{e11}). Note that equilibrium for the environment at temperature $1/\beta$ implies $\gamma(-\omega_0)=\eexp{-\beta\omega_0}\gamma(\omega_0)$. In the following we denote $\Gamma_e=\gamma(-\omega_0)$ (excitation rate described by $S_+$), $\Gamma_r=\gamma(\omega_0)$ (relaxation rate described by $S_-$), $2\Gamma_\phi=\gamma(0)$ (dephasing rate described by $S_z$). The conventional electron relaxation times are $\frac{1}{T_1}=\Gamma_e+\Gamma_r,\, \frac{1}{T_2}=\Gamma_\phi+\frac{1}{2T_1}$.

 In the following we reorder the density matrix $\rho_{ij}$ as a vector with 16 components and then Lindblad's equation becomes a matrix equation \cite{martin} in the super-space 16$\times$16 of the form $\frac{d\rho}{dt}=L\cdot \rho$. To evaluate correlation functions we employ the regression theorem \cite{lax,gardiner}, that follows from an assumption that at the initial time the system and environment density matrices are decoupled. In fact, the Markovian assumption, needed for deriving the Lindblad equation, guarantees that the system and environment equilibrate fast and we can choose the equilibrium system density matrix $\rho_{st}$ as the initial state.

 For the STM-ENDOR experiment we are interested in the spin-spin correlation. In order to evaluate Fourier transforms we need to convert the $t<0$ time integration into a $t>0$ integration by using stationarity of the correlation. The result has the form
 \beq{14}
C_{-+}(\nu)&=&\int_{-\infty}^\infty \langle S_-(t)S_+(0)\rangle\eexp{i\nu t}dt\nonumber\\
&=&C_{1a}+C_{2a}+C_{1b}+C_{2b}\nonumber\\
C_{1a}&=&\int_0^\infty  Tr[\eexp{i{\cal H}t}S_{a-}\eexp{-i{\cal H}t} S_{a+}\rho_{st}\rho_{env}]\eexp{i(-\omega +\nu)t}dt\nonumber\\ &=&Tr[(S_{a-}\otimes U)\frac{1}{s-L}(S_{a+}\otimes U)\rho_{st}]_{s\rightarrow i(\omega-\nu)}\nonumber\\
C_{2a}&=&\int_0^\infty Tr[S_{a-}\eexp{i{\cal H}t}S_{a+}\eexp{-i{\cal H}t}\rho_{st}\rho_{env}]\eexp{i(\omega-\nu) t}dt\nonumber\\ &=&Tr[(S_{a+}\otimes U)\frac{1}{s-L}(U\otimes S_{a-}^T)\rho_{st}]_{s\rightarrow i(-\omega+\nu)}\nonumber\\
\eeq
and $C_{1b},C_{2b}$ are obtained from $C_{1a},C_{2a}$ by $a\rightarrow b,\,\omega\rightarrow -\omega$; $\rho_{env}$ is the environment density matrix and $U$ is a $4\times 4$ unit matrix. Finally the $C_{-+}(\nu)$ is obtained from $C_{+-}(\nu)$ by replacing $+\leftrightarrow -$ and $\omega\rightarrow -\omega$. We have solved for the correlations using Mathematica, noting that the inverse matrix $\frac{1}{s-L}$ could be found analytically. The solutions are shown, with parameters relevant to the Si data, in Fig. 6 of the main text.

It is also useful to solve for the eigenvalues of Eq. (\ref{e11}) (without $\tilde{\cal H}_{SE,rot}$)  and identify the ESR resonances. In particular we find a mode, in the laboratory frame at
\beq{15}
\omega_{ESR}=&&\omega_0+\half\sqrt{(\omega_n+\half a-\omega)^2+h^2}\nonumber\\ &&-\half\sqrt{(\omega_n-\half a+\omega)^2+h^2}+\omega
\eeq
At $\omega=\half a$ this yields the original ESR frequency $\omega_{ESR}=\omega_0+\half a$, hence negative ENDOR is weakened and we expect a peak at $\omega=\half a$, independent of $h$, as indeed seen in the data Fig. 2b,c and in the solutions Fig. 6  of the main text. The maxima of negative ENDOR, i.e. dips in the data, do shift with $h$.

 \end{document}